\documentclass[
  aps,prl,reprint,
  superscriptaddress,
  nofootinbib,
  longbibliography,
  nobalancelastpage
]{revtex4-2}
\usepackage[T1]{fontenc}
\usepackage[utf8]{inputenc}
\usepackage{xparse}
\usepackage{etoolbox}

\makeatletter
\newenvironment{plainwidetext}{%
  \par\ignorespaces
  \onecolumngrid
  \vspace{0\baselineskip}
}{%
  \par
  \vspace{0\baselineskip}
  \twocolumngrid\global\@ignoretrue
  \@endpetrue
}
\makeatother

\usepackage{amsmath,amssymb,bm}
\usepackage{graphicx}
\usepackage[colorlinks=true,allcolors=blue]{hyperref}
\newcounter{prlapp}

\newcommand{\prlappendix}[2]{%
  \par\vspace{1.0em}%
  \refstepcounter{prlapp}%
  \setcounter{equation}{0}%
  \renewcommand{\theequation}{\Alph{prlapp}\arabic{equation}}%
  \noindent\textit{Appendix \Alph{prlapp}: #1}---\label{#2}\ignorespaces%
}

\newcommand{\prlsubheading}[1]{%
  \noindent\textit{#1}---\ignorespaces%
}
\begin{document}
\title{Spontaneous patterning of cell size on curved surfaces}

\author{Yuan He}
\affiliation{School of Materials Science and Engineering, Zhejiang University, Hangzhou, Zhejiang 310027, China}
\affiliation{Department of Materials Science and Engineering, School of Engineering, Westlake University, Hangzhou, Zhejiang 310030, China}

\author{Shi-Lei Xue}
\thanks{Corresponding author: xueshilei@westlake.edu.cn}
\affiliation{Department of Materials Science and Engineering, School of Engineering, Westlake University, Hangzhou, Zhejiang 310030, China}

\begin{abstract}
Tissue surfaces exhibit complex curvature during embryogenesis and oncogenesis. Evidence shows that cells can actively sense curvature to regulate behavior and fate, yet the underlying mechanism remains unclear. Here, we develop a vertex model for arbitrary curved surfaces and uncover spontaneous cell size patterning on ellipsoidal surfaces: cells in high-curvature regions are consistently larger than those in low-curvature regions. This non-uniformity arises from a mechanical competition encoded in Riemannian geometry: positive Gaussian curvature reduces the perimeter-to-area ratio of polygonal cells, relaxing cell-edge tension in high-curvature regions, which is compensated by area expansion to maintain global force balance. This area pattern is robust against variations in model parameters and matches observations in biological systems. The perimeter pattern, in contrast, is governed by competition between the intrinsic geometric tendency and the deformation required by force balance, and undergoes reversal beyond a critical shape index. Together, these findings establish self-organized spatial variations in cell size as a potential physical mechanism for curvature sensing.
\end{abstract}

\maketitle

\emph{Introduction}---Tissue and organ surfaces exhibit complex topographies during key life processes such as embryogenesis~\cite{liu2024morphogenesis,hirashima2024erk,okuda2018strain,alber2025model,xue2025mechanochemical,collinet2021programmed,schamberger2023curvature}  and oncogenesis~\cite{lee2016interfacial,messal2019tissue,wirtz2011physics}. Surface curvature has been recognized as a global geometric boundary condition that influences tissue organization, drives spontaneous cell motions~\cite{xi2017emergent,brandstatter2023curvature,tang2022collective}, and gives rise to fascinating phenomena such as coherent cell rotation~\cite{glentis2022emergence,tan2024emergent,happel2024coordinated}. Importantly, emerging evidence now reveals that cells can actively sense surface curvature to locally modulate biochemical signals (e.g., Shh, BMP4, YAP/TAZ) and mechanical cues (e.g., actomyosin tension), which in turn direct cell behaviors such as constriction~\cite{xue2025mechanochemical}, extrusion~\cite{krueger2025epithelial}, migration~\cite{pieuchot2018curvotaxis}, and ultimately determine cell fate such as differentiation or stemness maintenance~\cite{xue2025mechanochemical,shyer2015bending,yavitt2023situ}. However, the mechanism by which tissue actively senses and responds to curvature remains largely unknown.

Among potential candidates, cell size provides a fundamental readout that directs epithelial tissue adaptation to environmental changes. On one hand, fluctuations in cell size modulate chemical concentrations and intracellular crowdedness, thereby regulating signaling activity and downstream cell behaviors~\cite{li2021volumetric,delarue2018mtorc1}. On the other hand, when cell size is limited, the nucleus becomes compressed by the cell cortex or cytoskeleton. This mechanical constraint affects YAP nucleocytoplasmic transport~\cite{koushki2023nuclear}, chromatin condensation, and DNA synthesis~\cite{kalukula2022mechanics}, ultimately impacting cell fate decisions~\cite{pentinmikko2022cellular}. These findings naturally lead to the question: do tissues actively sense substrate curvature also through the size of single cells?

To address this, we propose a vertex model for confluent tissues on curved substrates, in which cells conform completely to the underlying surface, and their edges follow geodesic paths between vertices. Surprisingly, the model reveals the spontaneous emergence of spatial non-uniformity in cell area and perimeter on non-spherical substrates, a phenomenon absent in tissue models on flat or uniformly curved surfaces~\cite{de2025epithelial,sussman2020interplay,hernandez2023finite}. This prediction provides a plausible explanation for the spatial variations in cell size observed in many epithelial tissues, and suggests that spontaneous cell size regulation may serve as a potential mechanism for curvature sensing.

\par\vspace{0.1\baselineskip}
\emph{Vertex model on curved manifolds}---We extend the planar~\cite{farhadifar2007influence} and spherical~\cite{de2025epithelial} vertex models to arbitrary tissue geometries by allowing cells to fully conform to the underlying surface and by defining cell edges as geodesic curves between neighboring vertices, rather than Euclidean straight lines (see End Matter, Appendix A). Our formulation thus incorporates the geometric constraints imposed by the curvature of tissue architecture. 

The dimensionless potential energy is \(e = \sum_{i=1}^{N}\frac{1}{2}\left[k_A(a_i-1)^2+(p_i-p_0)^2\right],\) where \(a_i\) and \(p_i\) are the area and perimeter of cell \(i\), respectively, and
\(k_A\) and \(p_0\) are the area stiffness and target shape index~\cite{bi2015density,alt2017vertex}. Importantly, on a curved surface, both \(a_i\) and \(p_i\) become functions of the local Gaussian curvature in Riemann normal coordinates (see End Matter, Appendix~A for details). For a regular \(n\)-polygonal cell, the area--perimeter relation depends
on the local Gaussian curvature \(\kappa\) as
\begin{equation}
p_i = c_n\sqrt{a_i}\left(1-\alpha_n\kappa a_i\right),
\label{eq:polygon_relation}
\end{equation}
\noindent with \(c_n=2\sqrt{n\tan(\pi/n)}\) and  \(\alpha_{n} = \left\lbrack \frac{5}{36}\cos^{2}\left( \frac{\pi}{n} \right)-\frac{1}{72} \right\rbrack/(n\sin{\frac{\pi}{n}\cos{\frac{\pi}{n})}}\) (see Supplemental Material (SM)~\cite{SM} Sec. A for more details). The shape index then becomes
\begin{equation}
\mathrm{SI}=p_i/\sqrt{a_i}=c_n(1-\alpha_n\kappa a_i).
\end{equation}
This relation indicates that, at fixed cell area, surface curvature reduces the perimeter of a polygonal cell. This curvature dependence of single-cell morphometric parameters implies that cell size is intrinsically curvature-sensitive. Motivated by this insight, we employ vertex simulations to extend our analysis to confluent tissues with random cell shapes. To account for intercellular mechanical interactions,  we focus on the elastic regime with \(p_0<p_0^*\), where \(p_0^*\) marks the jamming transition~\cite{bi2015density}. 

Prolate ellipsoidal geometries are prevalent in developmental biology, appearing in systems such as developing embryoes~\cite{huang2020embryonic,gehrels2023curvature}, ovarian follicles~\cite{jia2016automatic,he2010tissue}, and various organoids~\cite{gjorevski2022tissue}. Here, as a prototypical example, we implement our vertex model on such an ellipsoidal surface (see SM~\cite{SM} Sec.~B for parameterization). We set the total area of the ellipsoid to $N$, such that the mean cell area is normalized to unity, i.e., $\langle a_i\rangle = 1$. Starting from a Voronoi tessellation on a prolate ellipsoid, we first minimize the energy at $p_0=3.75$ and then systematically decrease the target shape index to probe the ground-state configurations of the tissue (see SM~\cite{SM} Sec.~C for more details). 

\begin{figure}[t]
    \centering
    \includegraphics[width=\columnwidth]{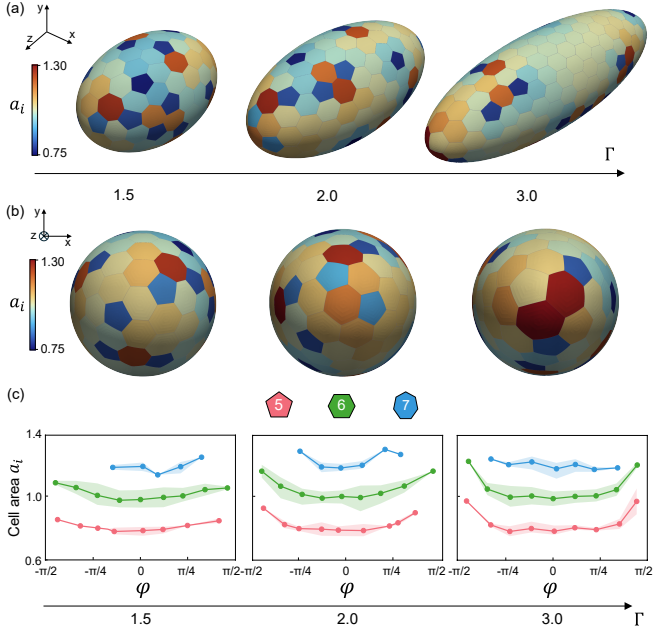}
    \caption{Spontaneous patterning of cell area on an ellipsoidal surface.
(a,b) Representative ground-state configurations showing that cells near the poles are larger than those near the equator. This area disparity increases with the aspect ratio \(\Gamma\) (minor axis fixed at 3; major axis varied). Color bar: cell area.
(c) Spatial distribution of cell area along the meridional angle \(\varphi\) for different cell types. Colors denote cell polygon types: red for pentagons, green for hexagons, blue for heptagons (see SM~\cite{SM} Sec.~A for detailed morphometric analysis). Markers indicate average values, and shaded regions represent the 95\% confidence interval.
}

    \label{fig:area_patterning}
\end{figure}
\par\vspace{0.1\baselineskip}
\emph{Cell area patterning}---We observe that cells in the polar region are consistently larger than those in the equatorial region [Fig.~1(a,b)]. Our quantification further shows that this spatial
non-uniformity becomes more pronounced as the aspect ratio \(\Gamma\) increases,
and is evident across pentagonal, hexagonal, and heptagonal cell types
[Fig.~1(c)]. The intrinsic geometric principle [Eq.~(1)] provides mechanistic insight into single-cell curvature sensitivity, yet it alone cannot explain why polar cells become larger than equatorial ones in a confluent tissue. This limitation points to the importance of intercellular mechanical interactions, which we hypothesize to be the origin of the spatial non-uniformity in cell area. Specifically, on a non-spherical surface, curvature variations are expected to
induce cell size variations [Eq.~(1)], and the resulting shape incompatibility
would in turn generate mechanical forces that deform the cells. To investigate
this hypothesis, we next analyze the mechanical stresses and deformation within
the tissue.
\begin{figure}[t]
    \centering
    \includegraphics[width=\columnwidth]{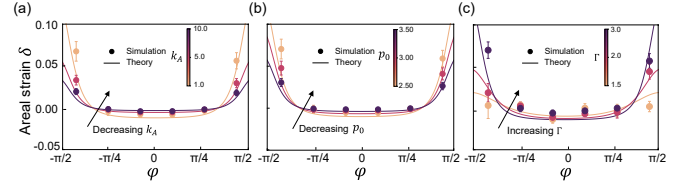}
    \caption{Areal strain distributions.
The areal strain \(\delta\) is plotted against the meridional angle \(\varphi\), while varying
(a) \(k_A\) at fixed \((p_0,\Gamma)=(2.5,3)\),
(b) \(p_0\) at fixed \((k_A,\Gamma)=(1,3)\), and
(c) \(\Gamma\) at fixed \((k_A,p_0)=(1,2.5)\).
Symbols denote simulation data (mean \(\pm\) SEM, sample number is 6), and solid lines represent theoretical predictions. The corresponding raw simulation data are shown in Fig.~S7~\cite{SM}.
}
    \label{fig:area_patterning}
\end{figure}

We coarse-grain the discrete system into a continuum elastic membrane. The tissue surface is parameterized by the meridional arclength \(s\) and the azimuthal angle \(\theta\). Let \(\lambda_{\theta}\) and \(\lambda_s\) respectively be the circumferential and meridional stretch ratios. The area change (or areal strain) of cells can be evaluated as \(\delta(s)=J-1\), with \(J=\lambda_{\theta}\lambda_s\) the ratio of current area to initial area. Adopting a mean-field approximation, the tissue free energy density takes the form
\begin{equation}
e(s)=\frac{k_A}{2}(J-1)^2+\frac{1}{2}\left[c_n\sqrt{J}\left(1-\alpha_n\kappa J\right)-p_0\right]^2.
\label{eq:free_energy_density}
\end{equation}
From this, the in-plane tissue tensions \(\sigma_s\) and \(\sigma_\theta\) follow as \((1/\lambda_\theta)(\partial e/\partial\lambda_s)\) and \((1/\lambda_s)(\partial e/\partial\lambda_\theta)\).
Meanwhile, force equilibrium dictates that the principal stresses \(\sigma_s\) and \(\sigma_\theta\) must be uniform across the ellipsoidal surface, i.e. \(\sigma_s=\sigma_\theta=const.\), so that all cells sustain the same stresses (see SM~Sec. D and F for details).
For small areal strain \(\delta\) [Fig.~1(c)] and small parameter \(\alpha_n\), these stresses can be linearized as
\begin{equation}
\sigma_s=\sigma_\theta=K(s)\delta+T(s)=const.,
\end{equation}
\noindent which naturally partitions the tissue stresses into two contributions: \(K(s)\delta\) is associated with cell area changes (with \(K\) the equivalent area stiffness), and \(T(s)\) represents the tension arising from cell-edge contractility. We have
\begin{subequations}
\label{eq:linear_stress}
\begin{align}
K(s) &= k_A + \frac{c_np_0}{4}, \\
T(s) &= -\frac{c_{n}\alpha_{n}}{2}\left(4c_{n}-3p_{0}  \right)\kappa(s)
   + \frac{c_{n}}{2}\left( c_{n} - p_{0} \right).
\end{align}
\end{subequations}

\noindent While the area stiffness \(K\) remains unaffected by local curvature [Eq. (5a)], the edge tension \(T\) decreases with increasing curvature [Eq. (5b)] as a mechanical consequence of the cell area-perimeter relation [Eq. (1)]: on high-curvature surfaces, cell perimeter shortens, leading to a relaxation of cell-edge tension. Eqs. (4) and (5) thus provide a mechanistic resolution to the origin of the spatial non-uniformity in cell area: the cell-edge tension \(T\) is lower in the polar regions, where the cell perimeter is shortened by curvature; to compensate for this reduction while maintaining global stress uniformity [Eq. (4)], polar cells undergo greater stretching deformation, which in turn enlarges their areas.

The force balance condition, together with the area constraint \(\int_{S}\delta(s)\,\mathrm{d}a_0 = 0 \), with
\(\mathrm{d}a_0\) the initial area element, yields the  analytical formula for the areal strain \(\delta\) (see SM~\cite{SM} Sec. E for more details):
\begin{equation}
\frac{\delta(s)}{\kappa(s)-\langle\kappa\rangle}
= \frac{2c_n\alpha_n(4c_n-3p_0)}{4k_A+c_np_0}
= \gamma(k_A,p_0).
\label{eq:area_curvature_relation}
\end{equation}

\noindent With \(n = 6\) for the dominant hexagonal cells, this mean-field model successfully captures the spatial patterning of cell area under various parameter settings (Fig.~2), revealing that this patterning is regulated by both cell area elasticity \(k_A\) [Fig.~2(a)] and the target shape index \(p_0\) [Fig.~2(b)], and becomes increasingly pronounced as the aspect ratio \(\Gamma\) increases [Fig.~2(c)]. By rescaling the simulation results of the areal strain \(\delta\) by the composite parameter \(\gamma(k_A,p_0)\), we find that the rescaled values across various \(k_A\) and \(p_0\) settings collapse onto a single master curve [Fig.~3(a)], validating the accuracy of the analytical formula. Notably, Eq.~(6) reveals a key insight: the polar-equatorial patterning is robust against variations in \(k_A\) and \(p_0\). Indeed, since \(\gamma(k_A,p_0)\) remains positive as long as \(p_0<c_n\), cells in the polar regions \((\kappa>\langle\kappa\rangle)\) are always stretched while those in the equatorial region are compressed (Fig.~3).

\begin{figure}[t]
    \centering
    \includegraphics[width=\columnwidth]{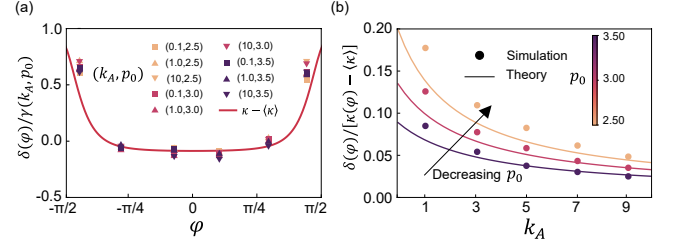}
    \caption{Comparison of vertex simulations and mean-field theory.
(a) Areal strain \(\delta\) rescaled by \(\gamma(k_A,p_0)\) collapses onto
the curvature deviation \(\kappa(\varphi)-\langle\kappa\rangle\) for different
parameter settings. Symbols denote mean values of simulation data, and the
solid line represents the master curve suggested by the mean-field theory.
(b) Global average of \(\delta(\varphi)/[\kappa(\varphi)-\langle\kappa\rangle]\) for different \(k_A\) and \(p_0\). The ratio increases as \(k_A\) and \(p_0\) decrease. Dots denote simulation results and
solid lines show theoretical predictions. The aspect ratio \(\Gamma=3\).
}
    \label{fig:area_patterning}
\end{figure}

\par\vspace{0.1\baselineskip}
\emph{Cell perimeter patterning}---The same mean-field analysis yields an analytical relation between cell perimeter and local curvature:
\begin{equation}
\frac{p(s)-\langle p\rangle}{\kappa(s)-\langle\kappa\rangle}
= c_n\left[\frac{c_n\alpha_n(4c_n-3p_0)}{4k_A+c_np_0}-\alpha_n\right]
= \chi(k_A,p_0),
\label{eq:perimeter_curvature_relation}
\end{equation}
where \(\chi(k_A,p_0)\) quantifies the sensitivity of perimeter to curvature variations. Surprisingly, Eq.~(7) implies that the robustness of polar-equatorial patterning may not hold for cell perimeter, whose pattern can be reversed (i.e. polar cells exhibit shorter perimeters than equatorial cells) once the target shape index exceeds the critical value
\begin{equation}
p_0^f(n)=c_n-\frac{k_A}{c_n}.
\label{eq:critical_shape_index}
\end{equation}

This reversal arises from the competition between two opposing effects in high-curvature regions: the intrinsic geometric tendency toward shorter perimeters [Eq.~(1)] and the stretching deformation required by force balance [Eq.~(4)]. It occurs for sufficiently large \(k_A\) and \(p_0\) [Eq.~(7)], where cell area variation is suppressed: larger \(p_0\) reduces tension, allowing cells to maintain their preferred areas, while larger \(k_A\) penalizes area deviations. In this limit, Eq.~(1) reduces to \(p(s)\approx c_n(1-\alpha_n \kappa)\), yielding an anti-correlation between cell perimeter and local curvature:
\begin{equation}
p(s)-\langle p\rangle \approx -c_n \alpha_n \left[\kappa(s)-\langle\kappa\rangle\right].
\label{eq:perimeter}
\end{equation}

Our vertex simulations confirm these predictions. Increasing area stiffness \(k_A\) or target shape index \(p_0\) shifts the peak cell perimeter from the poles toward the equatorial region  [Fig.~4(a,b)]. Simulation data for various \(k_A\) and \(p_0\) collapse onto the master curve given by Eq.~(7) [Fig.~4(c)], and this collapse also holds for ellipsoidal tissues with different aspect ratios \(\Gamma\) (Fig.~S6~\cite{SM}). To further test the mean-field model, we construct a phase diagram of the polar-equatorial contrast in cell perimeter [Fig.~4(d)], which clearly delineates two distinct phases: for small \(k_A\) and \(p_0\), polar cells exhibit larger perimeters than equatorial cells; for large \(k_A\) and \(p_0\), this pattern is reversed. The phase boundary is accurately predicted by Eq.~(8), confirming the quantitative agreement between theory and simulation.

\begin{figure}[t]
    \centering
    \includegraphics[width=\columnwidth]{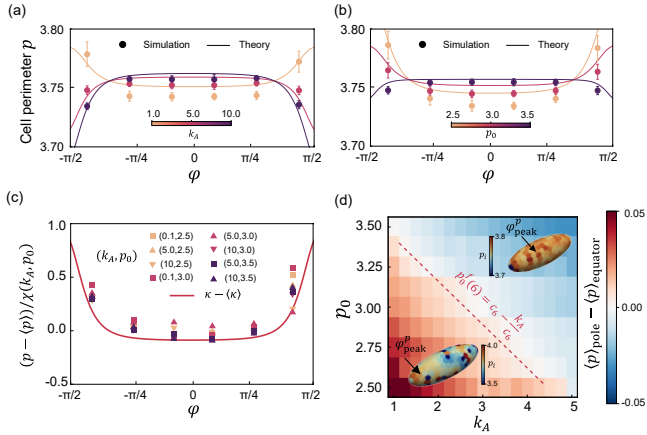}
    \caption{Patterning of cell perimeter on ellipsoidal surface. The distribution of
cell perimeter \(p\) along the meridional angle \(\varphi\) for (a) various
\(k_A\) at \(p_0=3\) and (b) various \(p_0\) at \(k_A=2\). Symbols denote
simulation data (mean \(\pm\) SEM, sample number is 6), and solid lines
represent mean-field predictions. (c) Rescaled simulation results (symbols) for
the perimeter deviation \([p(s)-\langle p\rangle]/\chi(k_A,p_0)\) collapse onto
the predicted master curve \(\kappa(\varphi)-\langle\kappa\rangle\) (red line)
for all tested \(k_A\) and \(p_0\). (d) Phase diagram of pole--equator contrast
of cell perimeter as a function of \(k_A\) and \(p_0\), with the color map
showing \(\langle p\rangle_{\mathrm{pole}}-\langle p\rangle_{\mathrm{equator}}\) and the red dashed line denoting the predicted transition threshold. The pole and
equator regions are defined by \(\left| \varphi \right|\in [\pi/3,\pi/2]\) and \(\left| \varphi \right|\in [0,\pi/6]\), respectively.
Inset: representative configurations colored by cell perimeter \(p_i\). The
aspect ratio \(\Gamma=3\). The corresponding raw simulation data are shown in Fig.~S8~\cite{SM}.
}
    \label{fig:area_patterning}
\end{figure}

\par\vspace{0.1\baselineskip}
\emph{Discussion}---The polygonal geometry of the apical/basal surfaces of epithelial cells is intrinsically sensitive to Gaussian curvature, characterized by shorter cell perimeters on positively curved surfaces than on flat ones. This sensitivity, when combined with spatially varying curvature (ubiquitous for tissue surfaces), creates geometric incompatibility and drives intercellular mechanical interactions, causing non-uniform distributions of cell areas and perimeters to emerge on curved surfaces. This prediction is confirmed in the epithelial tissue of \textit{Drosophila} embryos, whose apical cell areas show a clear dependence on local tissue curvature and are well described by the analytic relation [Eq. (6)] (see End Matter, Appendix B). Our findings thus suggest a potential curvature-sensing mechanism: cells first respond to curvature by changing their size, which in turn generate intercellular mechanical interactions or alter local biochemical concentrations, thereby activating downstream signaling pathways. 


Our vertex model can be readily applied to other geometric configurations, such as tori, Gaussian bumps and sinusoidal undulations ~\cite{luciano2021cell,mobasseri2019patterning}. It can also be used to analyze the influence of surface curvature on tissue rheology, particularly solid-fluid transitions~\cite{de2025epithelial}. Moreover, additional mechanical elements of epithelial cells, such as active fluctuations in line tension~\cite{rizzi2026universal,kim2021embryonic,claussen2024geometric,curran2017myosin, krajnc2020solid}, active bulk stress~\cite{yu2025feedback,lin2023structure,rozman2024cell}, or mechano-chemical feedback~\cite{maroudas2024confinement,sknepnek2023generating,boocock2021theory,krajnc2021active}, can be integrated into the model to study more realistic morphogenetic processes and their associated curvature sensing mechanisms. Finally, a fully 3D description of cell shapes that incorporates apical, basal, and lateral cell surfaces~\cite{xue2025mechanochemical,hannezo2014theory,drozdowski2026cell,rozman2020collective,merkel2018geometrically} would better recapitulate the complex self-adaptation of cell morphology (including cell height and skewness) to curved surfaces.

\section*{ACKNOWLEDGMENTS}

The authors acknowledge support from the National Natural Science Foundation of China
(Grant No. 12402073).

\let\oldbibitem\bibitem
\RenewDocumentCommand{\bibitem}{o m}{%
  \ifstrequal{#2}{merkel2018geometrically}{\newpage}{}%
  \IfNoValueTF{#1}
    {\oldbibitem{#2}}
    {\oldbibitem[#1]{#2}}%
}

\bibliography{refs}
\clearpage

\begin{plainwidetext}
\begin{center}
{\large\bfseries End Matter}
\end{center}
\end{plainwidetext}

\prlappendix{Vertex model on arbitrary curved surfaces}{app:vertex_model}
In the vertex model, the changes in cell shape and arrangement are achieved through the movements of polygon vertices, which are driven by the potential forces applied to them. The potential force applied to vertex \(\alpha\) is \(\bm{F}_{\alpha} = - \sum_{i\in\mathcal C(\alpha)}^{}{\partial e_{i}/\partial\bm{r}_{\alpha}}\), with \(\bm{r}_{\alpha}\) the spatial position of vertex \(\alpha\) and \(\mathcal C(\alpha)\) denotes the set of cells sharing vertex \(\alpha\). The vertex motion follows overdamped dynamics, in which this potential force is balanced by a viscous drag
\begin{equation}\label{eq:B1}
    \eta \frac{d\bm r_\alpha}{dt}
= \bm F_\alpha,
\end{equation}
where \(\eta\) is the viscous coefficient.

In this section, we explicitly calculate the expressions used to compute forces on arbitrary curved surfaces. As in the previous work in flat space \cite{bi2015density}, gradients of the energy with respect to vertex positions \(\bm{r}_\alpha\) decompose by the chain rule into 
\begin{equation}
\begin{aligned}
\bm F_\alpha
&=
-\sum_{i\in\mathcal C(\alpha)}
\left[
\frac{\partial e_i}{\partial a_i}
\frac{\partial a_i}{\partial \bm r_\alpha}
+
\frac{\partial e_i}{\partial p_i}
\frac{\partial p_i}{\partial \bm r_\alpha}
\right] \\
&=
-\sum_{i\in\mathcal C(\alpha)}
\left[
k_A(a_i-1)
\frac{\partial a_i}{\partial \bm r_\alpha}
+
(p_i-p_0)
\frac{\partial p_i}{\partial \bm r_\alpha}
\right].
\end{aligned}
\end{equation}
\noindent Thus, the key geometric ingredients are: (a) the area and perimeter of geodesic polygons on the curved surface, (b) their derivatives with respect to vertex coordinates, and (c) the geodesic distances between neighboring vertices. In simulation, we evaluate (a) and (c) using the open-source libraries GeographicLib \cite{karney2013geodesics}, which computes geodesic lengths, endpoint tangent vectors, and polygonal areas on an ellipsoid. Below, we first derive local analytical approximations for (a), which are then used in the mean-field theory and in deriving the analytical expressions for (b). Our framework can be further extended to other smooth curved surfaces \cite{webb2025curvedspacesim}.

\prlsubheading{Area and perimeter of cells on curved surfaces} We first consider a regular \(n\)-polygonal cell \(D\) residing on a curved
surface with Gaussian curvature field \(\kappa\). We work in a Riemann normal-coordinate patch centered at the cell centroid \(x_0\). Since the cell size is much smaller than the characteristic radius of curvature of the tissue, we have
\(|\kappa(x_0)|R^2\ll 1\), where \(R\) is the circumradius of the cell. Under this local approximation, the cell area is given by (see SM~\cite{SM} Sec.~A for more details): 
\begin{equation}\label{eq:D6}
a
=
a_E
\left[
1
-
\frac{\kappa(x_0)}{12}R^2 f(n)
\right]
+
O(R^5),
\end{equation}
where
\(a_E
=
nR^2\sin\left(\frac{\pi}{n}\right)
\cos\left(\frac{\pi}{n}\right)\)
is the flat-space area of the regular \(n\)-polygon, and
\(f(n)
=
\frac{1}{3}
+
\frac{2}{3}\cos^2\left(\frac{\pi}{n}\right).\)

Similarly, the cell perimeter is (see SM~\cite{SM} Sec.~A for more details):
\begin{equation}
p
=
p_E
\left[
1
-
\frac{\kappa(x_0)}{6}
R^{2}\cos^{2}\left(\frac{\pi}{n}\right)
\right]
+
O(R^{4}),
\end{equation}
with
\(p_E
=
2nR\sin\left(\frac{\pi}{n}\right)\) the flat-space perimeter of the regular \(n\)-polygon. For notational simplicity, we write \(\kappa\equiv \kappa(x_0)\) in the following sections.

\prlsubheading{Derivative of cells on curved surfaces} The perimeter of a cell \(i\) is the sum of the lengths of its boundary geodesics,
\begin{equation}\label{eq:C1}
\ p_{i} = \sum_{(\alpha,\beta) \in i}^{}l_{\alpha\beta},
\end{equation}
\noindent where \(l_{\alpha\beta}\) is the geodesic length between vertices \(\bm{r}_{\alpha}\ \)and \(\bm{r}_{\beta}\). For an edge \((\alpha,\beta)\), let \(\bm{\tau}_{(\alpha,\beta),\alpha}, \bm{\tau}_{(\alpha,\beta),\beta}\)
denote the unit tangent vectors of this geodesic evaluated at the endpoints \(\bm{r}_{\alpha}\ \)and \(\bm{r}_{\beta}\), respectively. Then the first variations of the edge length satisfy
\begin{equation}\label{eq:C2}
 \frac{\partial l_{\alpha\beta}}{\partial\bm{r}_{\alpha}} = \bm{\tau}_{(\alpha,\beta),\alpha},\ \ \frac{\partial l_{\alpha\beta}}{\partial\bm{r}_{{\beta}}} = \bm{\tau}_{(\alpha,\beta),\beta}.
\end{equation}

\noindent The derivative of \(p_i\) with respect to the vertex position \(\bm{r}_{\alpha}\) is simply expressed as
\begin{equation}\label{eq:C3}
 \frac{\partial p_i}{\partial\bm{r}_{\alpha}} = \bm{\tau}_{(\alpha,\alpha + 1),\alpha} + \bm{\tau}_{(\alpha - 1,\alpha),\alpha}.
\end{equation}

\noindent Therefore, for the edge \((\alpha,\beta)\) shared by cells \(i\) and \(j\), the contribution to the force at vertex \(\alpha\) is 
\begin{equation}
    \left\lbrack \left( p_{i} - p_{0} \right) + \left( p_{j} - p_{0} \right) \right\rbrack\bm{\tau}_{(\alpha,\beta),\alpha},
\end{equation}
with the same magnitude as in flat space \cite{li2019mechanical}.


Based on the area expansion in Eq.~\eqref{eq:D6}, we derive the geometric derivative of the area with respect to the vertex positions. We first decompose the area into a Euclidean baseline and a curvature-dependent term:
\begin{equation}\label{eq:C4}
a(D) = a_{E}(D) - \frac{\kappa}{6} M_{2}(D) + \mathcal{O}\left( R^{5} \right),
\end{equation}
where \(a_E(D)\) is the Euclidean area of the polygon \(D\), and \(M_{2}(D) = \int_D \left| x \right|^{2}\, d^{2}x\) is the second moment of the domain about the origin \(x_0\).

Consider a counterclockwise polygon with vertices \(\bm{r}_1, \ldots, \bm{r}_n\), all measured from \(x_0\). Define
\[
\begin{aligned}
\bm{\Omega} &=
\begin{bmatrix}
0 & 1 \\
-1 & 0
\end{bmatrix},
\qquad
b_{\alpha} = \bm{r}_{\alpha}\cdot \bm{\Omega} \cdot \bm{r}_{\alpha+1}, \\
s_{\alpha}
&= \left| \bm{r}_{\alpha} \right|^{2}
+ \bm{r}_{\alpha} \cdot \bm{r}_{\alpha+1}
+ \left| \bm{r}_{\alpha+1} \right|^{2}.
\end{aligned}
\]
The Euclidean polygon area is \(a_{E}(D)=\frac{1}{2}\sum_{\alpha} b_{\alpha}\). To evaluate \(M_2(D)\), we decompose \(D\) into wedges \(\Delta_{\alpha}=\operatorname{conv}\{\mathbf{0},\bm{r}_{\alpha},\bm{r}_{\alpha+1}\}\).
A straightforward calculation gives \(\int_{\Delta_{\alpha}} \left| x \right|^{2} d^{2}x
= \frac{b_{\alpha}s_{\alpha}}{12}.\) Summing over all wedges, we obtain
\begin{equation}\label{eq:C7}
M_{2}(D) = \frac{1}{12}\sum_{\alpha=1}^{n} b_{\alpha}s_{\alpha}.
\end{equation}
Using
\begin{equation}\label{eq:C8}
\begin{aligned}
\frac{\partial a_{E}}{\partial \bm{r}_{\alpha}} &= \frac{1}{2} \bm{\Omega} \cdot \left( \bm{r}_{\alpha+1} - \bm{r}_{\alpha-1} \right), \\
\frac{\partial b_{\alpha}}{\partial \bm{r}_{\alpha}} &= \bm{\Omega}\cdot  \bm{r}_{\alpha+1}, \\
\frac{\partial s_{\alpha}}{\partial \bm{r}_{\alpha}} &= 2\bm{r}_{\alpha} + \bm{r}_{\alpha+1},
\end{aligned}
\end{equation}
we obtain the derivative of \(a\) with respect to the vertex position \(\bm{r}_{\alpha}\):
\begin{equation}\label{eq:C9}
\begin{aligned}
\frac{\partial a}{\partial \bm{r}_{\alpha}}
&=
\frac{1}{2} \bm{\Omega}\cdot 
\left( \bm{r}_{\alpha+1} - \bm{r}_{\alpha-1} \right) \\
&\quad
-\frac{\kappa}{72}\Big[
\bm{\Omega}\cdot  \bm{r}_{\alpha+1} s_{\alpha}
- \bm{\Omega} \cdot \bm{r}_{\alpha-1} s_{\alpha-1} \\
&\qquad
+ b_{\alpha}\left( 2\bm{r}_{\alpha} + \bm{r}_{\alpha+1} \right)
+ b_{\alpha-1}\left( 2\bm{r}_{\alpha} + \bm{r}_{\alpha-1} \right)
\Big]
.
\end{aligned}
\end{equation}

\noindent In the flat-space limit \(\kappa=0\), Eq.~\eqref{eq:C9} reduces to
\begin{equation}
\frac{\partial a}{\partial \bm{r}_{\alpha}}
=
\frac{1}{2} \bm{\Omega} \cdot \left( \bm{r}_{\alpha+1} - \bm{r}_{\alpha-1} \right),    
\end{equation}
\noindent which is the standard expression for the Euclidean polygon area gradient.

Together with the corresponding perimeter derivative, this result provides the geometric ingredients needed to construct the explicit overdamped equations of motion used in the simulations.





\begin{figure}[htbp]
    \centering
    \includegraphics[width=1\linewidth]{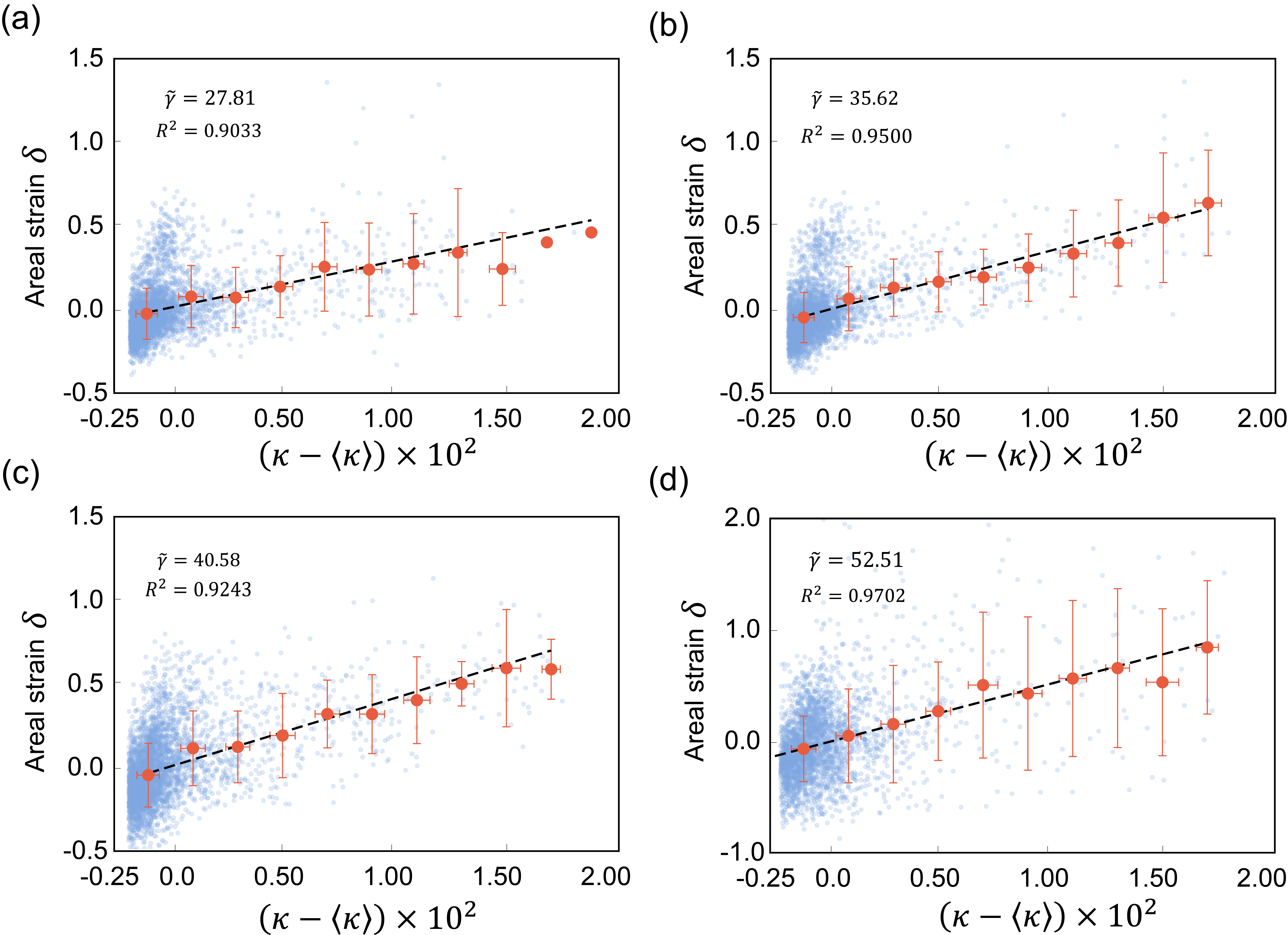}
    \caption{Apical cell areal strain as a function of Gaussian curvature for four \textit{Drosophila} embryos before ventral furrow formation. Blue dots represent single-cell measurements \cite{yang2025multicell,stern2022deconstructing}, and orange circles with error bars denote the binned averages and standard deviations. Dashed lines show linear fits to the binned data. The fitted slope \(\tilde\gamma\)  and coefficient of determination \(R^2\) are computed from the binned data in each panel.}
    \label{fig:S5}
\end{figure}

\prlappendix{Cell area--curvature relation in \textit{Drosophila} embryo}{app:drosophila}
To test the reasonability  of our model, in this section, we analyze the apical-surface data of \textit{Drosophila} embryos from Refs.~\cite{stern2022deconstructing,yang2025multicell}. We focus on the stage before ventral furrow formation, when the embryo surface remains well approximated by an ellipsoid. For each cell, we extract its apical area and Gaussian curvature, from which we compute the areal strain \(\delta\).
As shown in Fig.~\ref{fig:S5}, despite substantial scatter in the single-cell measurements, the binned averages increase with curvature, in agreement with our model prediction. Moreover, all four embryos exhibit a clear linear dependence of \(\delta\) on curvature. To reduce the influence of cell-to-cell fluctuations, we perform linear fits to the binned data, from which we obtain the slope \(\tilde\gamma\) and the coefficient of determination \(R^2\). The fitted slopes are consistently positive, ranging from \(27.8\) to \(52.0\). These high slopes indicate that tissue mechanics at this stage is primarily tension dominated, consistent with the physical picture proposed in recent work \cite{claussen2024geometric,brauns2024geometric}.

\end{document}